\documentstyle[11pt]{article}

\def\beq{\begin{eqnarray}}
\def\eeq{\end{eqnarray}}

\def\al{\alpha}
\def\be{\beta}

\def\ga{\gamma}

\def\vp{\varepsilon}
\def\ep{\epsilon}

\def\la{\lambda}
\def\na{\nabla}

\def\Ga{\Gamma}

\def\Si{\Sigma}

\begin{document}





\centerline{\bf A NOTE ON THE HEAT KERNEL METHOD}
\baselineskip=13pt
\centerline{\bf APPLIED TO FERMIONS}
\vspace*{0.37truein}
\centerline{G. DE BERREDO-PEIXOTO\footnotesize
\footnote{Electronic address: gpeixoto@phys.ualberta.ca}}
\baselineskip=12pt
\centerline{\footnotesize\it Department of Physics, University
of Alberta}
\baselineskip=10pt
\centerline{\footnotesize\it 412 Avadh Bhatia Physics Laboratory,
Edmonton, AB, T6G 2J1 Canada.}
\baselineskip=12pt
\centerline{\footnotesize\it Instituto de F\'{\i}sica, Universidade
de Bras\'{\i}lia}
\baselineskip=10pt
\centerline{\footnotesize\it Bras\'{\i}lia, DF, 70910-900
Brazil;}


\vskip 10mm

{\centerline{\large \it {\sl ABSTRACT}}}
\vskip 3mm

\begin{quotation}
The spectrum of the fermionic operators
depending on external fields is an important object in quantum field
theory. In this
paper we prove, using transition to the alternative basis for the
$\ga$-matrices, that this spectrum does not depend on the sign
of the fermion mass, up to a constant factor.
This assumption has been extensively used, but usually without proof.
As an illustration, we calculated the coincidence limit of
the coefficient $a_2(x,x^\prime)$ on the general background of metric,
vector and axial vector fields.
\end{quotation}

\vskip 10mm
\section{Introduction}	
\vspace*{-0.5pt}
\noindent
In four space-time dimensions, a relevant part of the spectrum
of a differential operator is the coincidence limit of the
coefficient\footnote{The coefficients $a_i$ are also called the
Hadamard-Minackshisundaram-Seeley-DeWitt coefficients} $a_2(x,x^\prime)$
in the
Schwinger-DeWitt expansion \cite{dewitt}. This coefficient defines the
divergent part of the effective action and is closely related to the
anomalies. The calculation of the differential operator spectrum is a
mathematical problem with important applications in theoretical physics; it
defines the relation between matter fields and quantum corrections to the
geometry (for general introduction, see \cite{birrell,book}).
Among various papers that investigate these physical
applications, we can cite \cite
{gold,kimu1,obu,nieh,bucsh,cogzer,satoshi,keon,elizalde}.  One can mention
that the papers \cite{gold,kimu1,obu,nieh,bucsh,cogzer,satoshi} consider
torsion,  which gives the general coupling between geometry and fermions
(see also \cite{book} and \cite{shapiro}). The calculation of the
determinant of a differential operator corresponding  to some particular
matter field is a necessary tool to investigate the effective action at the
1-loop level. In many cases, the 1-loop contribution contains the most
relevant information of the quantum effects in the low energy regime. The
1-loop part of the effective action can be written as (see \cite{book} )
\begin{eqnarray}
\Gamma ^{(1)}[\phi ]=\ep\,\frac{i}{2}\, {\rm ln} \, {\rm det}\, S_2[\phi ]=
\,\ep\,\frac{i}{2}\,\mbox{Tr}\,\mbox{ln}\, S_2[\phi ]  \label{det}
\end{eqnarray}
where $\phi $ is an arbitrary set of background fields (mean fields),
$\hat{H} = S_2[\phi ]$ is the bilinear functional operator $\delta ^2S[\phi
]/\delta\phi (x)\delta\phi (y)$ and $\ep=+1$ for bosonic fields and
$\ep=-1$ for fermionic fields. One can use the Schwinger-DeWitt method of
proper  time expansion to calculate the determinant (\ref{det}). The heat
kernel approach has been widely developed for many applications in Physics,
as well as a tool of mathematical nature (see, for example, \cite{avramidi}
and references therein). For instance, Goldthorpe \cite{gold} considered a
minimal second order operator in a Riemann-Cartan background, and
calculated the asymptotic expansion coefficient, $a_2$, considering
contributions from the skew-symmetric torsion.   Nieh and Yan \cite{nieh}
used  also the proper-time expansion technique to compute the coefficient,
but they considered in addition all the invariants from the generic torsion
as well as the conformal and axial anomalies. Obukhov \cite{obu} has
generalized the Goldthorpe's result, by the use of the Seeley's  method
\cite{seeley}, finding the corresponding anomalies.

The first-order Dirac operator includes negative eigenvalues in its spectrum
(see \cite{elizalde,elizalde2}), so that one has to use a supposition about
the evenness in the mass structure of the spectrum and to introduce the conjugated
operator in order to make the application of the Schwinger-DeWitt method
possible (see \cite{abdalla,elizalde,keon}). The main purpose of the
present paper is to prove the required assumption, which is universaly
accepted by the authors who calculate the spectrum of the Dirac operator.
The 1-loop effective action (\ref{det}) corresponding to the Dirac action,
\beq 
S_{1/2}= i\,\int d^4x\,\sqrt{-g}\,{\bar \psi}\, \left[ \,\ga^\mu \, D_\mu - im
\,\right]\,\psi\, , \label{acaotor-vec} 
\eeq 
is
\beq 
\Ga ^{(1)}[\phi ]=
-i\, {\rm Tr}\, ln\left\{\,\hat{H}[\phi ]\,\right\}\, ,
 \eeq
 where
 \beq
 \hat{H} = i\ga ^{\mu}D_{\mu}+ m\, ,
 \eeq
 and $g$ is the metric determinant.
 Here $D_{\mu}$ is the covariant derivative, which includes usual
 couplings\footnote{$A_{\mu}$ is the Abelian vector field, and $S_{\mu}$
 is an axial vector field, which can be related to torsion in the
 Riemann-Cartan framework. Here $\ga^5 = i\ga^0\ga^1\ga^2\ga^3$}$\,\,$:
 \beq
 D_{\mu}=\nabla _{\mu}-ieA_{\mu}+i\eta\ga ^5\, S_{\mu}\, ,
 \eeq
 where $e$ and $\eta$ are dimensionless coupling constants and $\na _{\mu}$
 is the covariant derivative in the curved background without torsion,
 as defined in \cite{shapiro}.

 This paper is organized as follows. In section 2, we review the basic
 concepts and calculations widely assumed in the spectral theory applied to
 the Dirac operator.
 As an illustration, we present the general result
 for the 1-loop divergence calculation in the Dirac action in the background
 Abelian vector, torsion and metric fields, including the superficial terms
 which are useful for the axial and trace anomaly calculation.
 In section 3, we demonstrate that a special assumption pointed
 out in section 2 is true.

\setcounter{footnote}{0}
\renewcommand{\thefootnote}{\alph{footnote}}

\section{Fermionic Operator coupled to External Fields}
\noindent
In order to use the Schwinger-DeWitt proper-time technique, the following
 trick is assumed and applied to the calculation of the 1-loop
 divergences in Dirac theories on curved backgrounds
 (see \cite{keon,elizalde,elizalde2}):
 \beq
 -i\, {\rm Tr}\, ln\left\{ i\ga ^\mu D_\mu + m \right\}
 = -\frac{i}{2}\, {\rm Tr}\, ln\left\{
 \ga ^{\mu}D_{\mu}\ga ^{\nu}D_{\nu}+m^2\right\} + \; C \;
 ; \label{anommult2}
 \eeq
where $C$, which is formally divergent, does not depend on the background
quantities, so it is not relevant for physical considerations\footnote{The
dimensional regularization scheme is adopted to deal with
the important divergent quantities.}$\,$ .
Basically, it arises here from the contribution of the term $C_0$ in
the following relation (which is required to prove
(\ref{anommult2}))\footnote{Additionaly, a reparametrization of
the (independent) fields,$(\psi\, ,\,\bar{\psi})\to (i\psi\, ,\,i\bar{\psi})$
is done in order to arrive at (\ref{anommult2}) from (\ref{m}), and,
as the Jacobian is constant, this procedure produces a constant quantity,
divergent, which is absorbed by $C$.} $\,$ :
\beq
 -i\, {\rm Tr}\, ln(i\ga ^{\mu}D_{\mu}+m)=
 -i\, {\rm Tr}\, ln(i\ga ^{\mu}D_{\mu}-m) + \, C_0\, .\label{m}
 \eeq
Here $C_0$ is divergent and constant (see (\ref{C0}) and (\ref{preres}) below).
By straightforward calculation, one can write the second order
 operator in the right hand side of (\ref{anommult2}) in the
 useful form
 \beq
 \Box +R_{\mu}\nabla ^{\mu} +\Pi\, ,\label{H-D2}
 \eeq
 where
 \beq
 \, R_{\mu} & = & -2ieA_{\mu}+2\eta\Si _{\mu\nu}\ga ^5S^{\nu}; \nonumber \\
 \Pi & = & -ie\nabla ^{\mu}A_{\mu}+i\eta\ga ^5\nabla ^{\mu}S_{\mu}
 -e^2A^{\mu}A_{\mu}+\eta ^2S^{\mu}S_{\mu}-\frac{i}{2}e\ga ^{\mu}\ga ^{\nu}
 F_{\mu\nu}+ \nonumber \\
 & + & \frac{i}{2}\eta\ga ^{\mu}\ga ^{\nu} \ga ^5 S_{\mu\nu}-
 2ie\eta\Si ^{\mu\nu}\ga ^5A_{\mu}S_{\nu}-\frac{1}{4}R+m^2\, ,
 \label{valoresRPi}
 \eeq
 with
 $$
 \frac{1}{2}\ga ^{\mu}\ga ^{\nu}[\nabla _{\mu}\; ,\nabla _{\nu}]=
 \frac{1}{8}\ga ^{\mu}\ga ^{\nu}\ga ^{\rho}\ga ^{\la}R_{\mu\nu\rho\la}=
 -\frac{1}{4}R\, ,
 $$
 $$
 F_{\mu\nu}=\nabla _{\mu}A_{\nu}-\nabla _{\nu}A_{\mu},\;\;\;\;
 S_{\mu\nu}=\nabla _{\mu}S_{\nu}-\nabla _{\nu}S_{\mu}\;\; {\rm and} \;\;\;\;
 \Si _{\mu\nu}=i/2\; [\ga _{\mu}\; ,\ga _{\nu}]\, .
 $$
 The object $R_{\mu\nu\al\be}$ is the curvature tensor, 
$R^\al\mbox{}_{\be\mu\nu}=\partial _\mu\Ga ^\al\mbox{}_{\be\nu}+
\Ga^\la\mbox{}_{\be\nu}\Ga ^\al\mbox{}_{\la\mu}-\,(\mu\leftrightarrow\nu )$
and $R=g^{\mu\nu}R^{\al}\mbox{}_{\mu\al\nu}$. One can find the resulting
expression for the corresponding 1-loop divergences, by direct use of
theSchwinger-DeWitt algorithm:
 
 \beq
 \Ga ^{(1)}_{{\rm div}} & = &
 \frac{\mu ^{(n-4)}}{\vp}\int d^nx\sqrt{-g}\;\left\{\, \frac{2}{3}e^2F_{\mu\nu}^2+
 \frac{2}{3}\eta ^2 S_{\mu\nu}^2-8m^2\eta ^2S_{\mu}S^{\mu}-\frac{1}{3}m^2R+2m^4+
 \right.\nonumber \\
 & + & \left.\frac{1}{72}R^2-
 \frac{1}{45}R_{\mu\nu}^2-\frac{7}{360}R_{\mu\nu\rho\la}^2
 -\frac{4}{3}\eta ^2\, \Box (S^{\mu}S_{\mu}) + \right.\nonumber \\
 & + & \left.\frac{4}{3}\nabla _{\mu}(S^{\nu}\nabla _{\nu}S^{\mu}-
 S^{\mu}\nabla _{\nu}S^{\nu}) -\frac{1}{30}\Box R\right\}\, . \label{result1}
 \eeq

\section{The Evenness in the Mass Structure of the
Operator $i\ga ^{\mu}D_{\mu}+m$}
\noindent
We are going to prove the relation (\ref{m}), and as a byproduct,  
the validity of eq. (\ref{anommult2}). For this purpose
 we shall find a basis of the $\ga$-matrices, in which the
 correctness of (\ref{anommult2}) becomes obvious.
 Following \cite{geyer}, let us consider the Dirac matrices basis,
 $(\Ga ^{\mu}\, ,\Ga ^4)$, defined by:
 \beq
 \Ga ^{\mu}=i\ga ^5\ga ^{\mu}\; ; \;\;\; \Ga ^4=i\ga ^5\; .
 \eeq
 Notice that these matrices satisfy the Clifford algebra:
 $$
 \left\{\Ga ^{\mu}\; ,\Ga ^{\nu}\right\}=2g^{\mu\nu}\;\;\;\; {\rm and}\;\;\;\;\;
 \Ga ^4\Ga ^4=-1\; .
 $$
 Consider the operator $\hat{F}=\Ga ^4\hat{D}=\Ga ^4(i\Ga
 ^{\mu}D_{\mu}+m)$.
 The expression $-i\, {\rm Tr}\, ln(\hat{F})$ is physically
 equivalent to $-i\, {\rm Tr}\, ln\, (i\ga ^{\mu}D_{\mu}+m)$,
 because this quantity must not depend
 on which basis we choose, and the difference is a constant factor which
 is not relevant. One can write then
 \beq
 -\frac{i}{2}\, {\rm Tr}\, ln(\hat{F}^2) = -i\, {\rm Tr}\, ln(\hat{F})=
 -i\, {\rm Tr}\, ln\,(i\Ga ^{\mu}D_{\mu}+m)-i\, {\rm Tr}\, ln\,\Ga ^4 .
 \label{C0}
\eeq

 One can find, by direct computation,
 \beq
 \Ga ^{\mu}D_{\mu}\Ga ^{\nu}D_{\nu}& = &
 \Ga ^{\mu}\Ga ^{\nu}\,
 (\nabla _{\mu}\nabla _{\nu}-ie\nabla _{\mu}A_{\nu}-ieA_{\nu}\nabla _{\mu}
 -ieA_{\mu}\nabla _{\nu} + \nonumber \\
 & + & \eta\Ga ^4\nabla _{\mu}S_{\nu}
 +\eta\Ga ^4S_{\nu}\nabla _{\mu}
 -e^2A_{\mu}A_{\nu}-ie\,\eta\Ga ^4A_{\mu}S_{\nu}+ \nonumber \\
 & + & ie\,\eta\Ga ^4A_{\nu}S_{\mu}
 -\eta\Ga ^4S_{\mu}\nabla _{\nu}+\eta ^2S_{\mu}S_{\nu}).
 \eeq
 Using this expression to write down $\hat{F}^2$, and turning back to the
 usual basis, we arrive at the surprising result:
 \beq
 \hat{F}^2 = \Box +R_{\mu}\nabla ^{\mu} +\Pi \, , \label{res}
 \eeq
 with $R_{\mu}$ and $\Pi$ exactly the same defined by (\ref{valoresRPi}).
 The conclusion
is that
 \beq
 \hat{F}^2 = \hat{H} = (i\ga ^{\mu}D_{\mu}+m)\cdot (i\ga ^{\mu}D_{\mu}-m) .
 \eeq
 Thus, we achieved that
 \beq
 -i\, {\rm Tr}\, ln\hat{F} & = & -i\, {\rm Tr}\, ln\, (i\ga ^{\mu}D_{\mu}+m) +
 C_0 = -\frac{i}{2}\, {\rm Tr}\, ln\hat{F}^2 = \nonumber \\
 & = & -\frac{i}{2}{\rm Tr}\, ln\left\{
 (i\ga ^{\mu}D_{\mu}+m)\cdot (i\ga ^{\mu}D_{\mu}-m)\right\}. \label{preres}
 \eeq
which proves the desired result, eq. (\ref{anommult2}).
By the virtue of the result
 (\ref{res}) it is clear that one does not need to repeat the whole
 calculation done in the previous basis  in section 2  and led to (\ref{result1}).
Indeed, all the algebra will be  the same and the new basis is as good as the
 original one, for both satisfy the same Clifford algebra.

\section{Conclusion}
\noindent
We have proved the assumption concerning the parity in the mass term,
 for the case of the most general Dirac operator in external fields.
 As an application, the divergent part of the fermion contribution to
 the vacuum effective action has been derived.

 \vskip 5mm
 \noindent {\bf Acknowledgments.} \vskip 3mm
I would like to thank Prof. Ilya L. Shapiro
 for suggesting me to study the alternative basis for fermions and for the
 fruitful discussions of the result. I am also grateful to Centro
Brasileiro de Pesquisas F\'{\i}sicas.
This work was done with support from CNPq, a Brazilian
Government institution that provides development in science and technology.


\begin{thebibliography}{99}
 \bibitem{dewitt}  B.S. DeWitt, {\it Dynamical theory of groups and fields} (Gordon and
 Breach, New York, 1965).

 \bibitem{birrell}  N.D. Birrell and P.C.W. Davies, {\it Quantum Fields in Curved Space,}
 Cambridge Univ. Press, Cambridge (1982).

 \bibitem{book}   I.L. Buchbinder, S.D. Odintsov and I.L. Shapiro, {\it Effective Action
 in Quantum Gravity, }IOP \ \ Publishing-Bristol (1992).

 \bibitem{gold} W.H. Goldthorpe, {\it Spectral geometry and SO(4) gravity in a
 Riemann-Cartan spacetime}, Nucl. Phys. {\bf B 170} (1980) 307-328.

 \bibitem{kimu1} T. Kimura, {\it Expansion coefficient of heat kernel of Laplacian
 operator in Riemann-Cartan space}, J. Phys. A: Math. Gen. {\bf 14}
 (1981) 329. \\
 T. Kimura, {\it Conformal and axial anomalies in Riemann-Cartan space},
 Prog. Theor. Phys. {\bf 66} (1981) 2011.

 \bibitem{obu} Yu.N. Obukhov, {\it Spectral geometry of the Riemann-Cartan
 spacetime}, Nucl. Phys. {\bf B 212} (1983) 237-254; \newline
 Yu.N. Obukhov, {\it Spectral geometry of the Riemann-Cartan spacetime
 and the axial anomaly}, Phys. Lett. {\bf B 108} (1982) 308-310;

 \bibitem{nieh} H.T. Nieh and M.L. Yan, {\it Quantized Dirac field in curved Riemann-Cartan
 background. I. Symmetry properties, Green's function}, Ann. of Phys. {\bf 138} (1982) 237-259.
 
 \bibitem{bucsh}  I.L. Buchbinder, S.D. Odintsov and I.L. Shapiro,
 {\it Nonsingular cosmological model with torsion induced by vacuum
 quantum effects}, Phys. Lett. {\bf B 162} (1985) 92.

 \bibitem{cogzer}  G. Cognola and S. Zerbini, {\it Seeley-DeWitt coefficients in a Riemann-Cartan
 space-time}, Phys. Lett. {\bf B 214} (1988) 70; \\
 {\it Heat kernel expansion in geometric fields}, Phys. Lett. {\bf B 195} (1889) 435.

 \bibitem{satoshi}  S. Yajima, {\it Evaluation of the heat kernel in Riemann-Cartan space}, Class.
 Quant. Grav. {\bf 13} (1996) 2423-2435.

 \bibitem{keon}  D.G.C. McKeon and C. Schubert, {\it Phys. Lett. }{\bf B 440} (1998) 101.

 \bibitem{elizalde}  E. Elizalde, {\it On the concept of determinant for the differential
 operators of Quantum Physics,} JHEP {\bf 07: 015} (1999).

 \bibitem{shapiro} I.L. Shapiro, {\it Physical aspects of
 the space-time torsion}. (to appear in Phys. Rep.) [hep-th/1013093].

 \bibitem{avramidi} I.G. Avramidi, {\it Covariant techniques for computation  of the heat kernel},
 Rev. Math. Phys.  {\bf 11} (1999) 947-980.

 \bibitem{seeley} R.T. Seeley, {\it Proc. Symp. Pure Math.}, Amer. Math. Soc.
 {\bf 10} (1967) 288.



 \bibitem{elizalde2} G. Cognola, E. Elizalde and S. Zerbini, {\it Dirac Functional
 Determinants in Terms of the Eta Invariant and the Noncommutative Residue},
 hep-th/9910038.

 \bibitem{abdalla} E. Abdalla, M.C.B. Abdalla and K.D. Rothe,
 {\it 2-Dimensional Quantum
 Field Theory}, World Scientific, Singapore (1991).

 \bibitem{geyer} B. Geyer, D. Gitman and I.L. Shapiro,
 {\it Path integral and pseudoclassical action for spinning particle in
 external electromagnetic and torsion fields}, Int. J. Mod. Phys.
 {\bf A 15} (2000) 3861-3876.
\end{thebibliography}
\end{document}